# Evaluation of *In Vivo* Subject-Specific Mechanical Modeling of the Optic Nerve Head for Robust Assessment of Ocular Mechanics

Soumaya Ouhsousou[1*], Lucy Q. Shen[2], Chhavi Saini[2], Amin Pourasghar[1], Mengyu Wang[3¶], John C. Brigham[1¶]

[1] Department of Civil and Environmental Engineering, University of Pittsburgh, Pittsburgh, PA, USA

[2] Department of Ophthalmology, Massachusetts Eye and Ear, Harvard Medical School, Boston, MA, USA

[3] Department of Ophthalmology, Harvard Ophthalmology AI Lab, Schepens Eye Research Institute of Mass Eye and Ear, Harvard Medical School, Boston, MA, USA

\* Corresponding author

E-mail: soo16@pitt.edu

¶ Equal contribution as co-senior authors.

## Abstract

To establish the tissue regions necessary to accurately represent the mechanics of the optic nerve head (ONH), imaging data of the ONH from 2 healthy subjects were used to create *in vivo* subject-specific eye mechanical models considering distinct properties for all major ocular tissues. Tests were performed to evaluate the effect of the material properties and the inclusion of these tissues on the mechanics of the lamina cribrosa (LC), retina, and optic nerve. Then, the LC mechanical response due to variations in intraocular and intracranial pressures was evaluated to validate the modeling approach. The sclera stiffness has the largest impact on the mechanics of the LC, retina, and optic nerve, while Bruch's membrane has a negligible effect on these tissues' response. The validation tests showed increased LC strain with increased pressure and highest strains in the inferior and temporal subregion, as seen in literature studies. Consequently, accurate ONH mechanical representation can be obtained by only including those tissue regions identified as necessary.

# Introduction

    Visual impairment is a common public health issue that has a profound impact on individuals' quality of life and impedes their ability to perform daily activities with ease (McKean-Cowdin et al., 2007, 2008). Visual impairments are often secondary to various eye diseases that have a range of effects on important tissue regions of the eye, including the retina, lamina cribrosa (LC), and the optic nerve (i.e., postlaminar tissue). For example, diseases such as age-related macular degeneration, induced by vitroretinal traction can severely impact the structure and function of the retina, leading to vision loss or impairment (Hani et al., 2021; Phillips et al., 2022). For instance, application of continuous pulsatile stretch on retinal pigment epithelium was found to increase the secretion of vascular endothelial growth factor (Seko et al., 1999), leading to pathologic alterations in retinal structure (Nishinaka et al., 2018). By a similar mechanical pathway, increased intraocular pressure (IOP) can result in biomechanical alterations of the LC, a mesh-like structure that allows the retinal ganglion cell axons to leave the eye. This mechanical alteration can lead to glaucoma through progressive degeneration of the retinal ganglion cells and their axons, and subsequent visual field loss. The normal functioning of the bundled nerve fibers in the optic nerve can also be disrupted by mechanically induced disorders, such as indirect traumatic optic neuropathy and compressive optic neuropathy (Y. Li et al., 2020; Rodriguez-Beato & De Jesus, 2023). Thus, understanding the biomechanics of these critical tissue regions of the eye, including their material properties, strain distribution, and stress patterns, is vital for understanding and ultimately preventing multiple pathological mechanisms that can lead to vision loss.

    Several experimental and computational studies have been conducted to explore the mechanics of the retina, LC, and the optic nerve. Experimental studies have employed a range of techniques, including *ex vivo* mechanical testing and various *ex vivo* imaging approaches to characterize the material properties, response to mechanical stimuli, and structural changes of these ocular tissues (Bellezza et al., 2003; K. Chen et al., 2010; K. Chen & Weiland, 2010; Sigal et al., 2009). For instance, (Bellezza et al., 2003) performed laser-induced IOP elevation along with *ex vivo* digital imaging with monkey eyes and evaluated the resulting structural changes in the LC, particularly the anterior LC position, as well as laminar thickness. Several studies also focused on *ex vivo* assessment of material properties of retina and the optic nerve through uniaxial tensile experiments with excised animal eye tissues (K. Chen et al., 2010; K. Chen & Weiland, 2010). Alternatively, studies have evaluated the mechanics of eye tissues using computational analysis (Karimi, Rahmati, et al., 2022; Muñoz-Sarmiento et al., 2019; Sigal et al., 2009). For example, (Sigal et al., 2009) employed computational mechanics with the finite element method and used geometries based on post mortem human eye imaging data to evaluate the strain distributions within the LC and retina under acute changes in IOP. Using the same finite element approach, more tissue regions of the eye, including optic nerve, dura mater, and choroid, were evaluated in a study conducted by (Muñoz-Sarmiento et al., 2019) to evaluate the effect of IOP and intracranial pressure (ICP) on strains within eye tissues based on orbital imaging data of a healthy subject. In contrast to experimental methods, these computational studies have the benefit of estimating many different biomechanical responses (e.g., displacement, strain, stress, etc.) subject to a wide range of loading conditions. However, there have been relatively large variations in the overall approaches used for such eye computational biomechanical modeling to-date.

    Several computational mechanical modeling approaches for the eye have been introduced, mainly using the finite element method (Eilaghi et al., 2010; Esposito et al., 2015; Kharmyssov et al., 2021; Roberts et al., 2010a; Sigal et al., 2009; Wang et al., 2016). These approaches varied significantly in terms of complexity (e.g., what tissue regions are included/considered) and subject specificity (e.g., whether using directly measured tissue dimensions or generic estimates). Scientists have constructed generic eye geometries with varying degrees of complexity, defining the tissue region geometries with simplified shapes and dimensions based on dimensions of the eye from literature data. For instance, (Eilaghi et al., 2010) constructed an axisymmetric eye finite element model that included tissue regions estimated with standard dimensions from the literature, including prelaminar tissue, sclera, LC, postlaminar tissue, and pia mater. Even more tissue regions were incorporated into a generic ONH

geometry and embedded into a corneoscleral sphere in work by (Wang et al., 2018), which included the annular ring, choroid, border tissue, and Bruch's membrane, in addition to the five tissue regions used in (Eilaghi et al., 2010). Alternatively, subject-specific eye geometries were also considered with different degrees of complexity. For instance, (Roberts et al., 2010a) created a simplified subject-specific eye finite element model from imaging data of monkey eyes that included only the sclera and the LC in the optic nerve head (ONH), which was embedded into a generic sphere to represent the eyeball. (Sigal et al., 2009) considered five tissue components for the ONH: prelaminar tissue, sclera, LC, pia mater, and optic nerve, to construct subject-specific finite element models of the ONH embedded into a generic sphere representing the eye globe, with tissue region geometry based on photographs from stained serial sections from 10 post-mortem human eyes. Even more tissue components were incorporated by (Kharmyssov et al., 2021), who added dura mater to their *in vivo* subject-specific ONH models based on spectral domain optical coherence tomography (OCT) scans and embedded into a generic sphere representing the eyeball.

The diversity of the mechanical modeling approaches for the eye has predictably resulted in variation in the biomechanical observations they produced. (Sigal et al., 2004) performed a direct comparison of three different mechanical finite element modeling approaches, distinguished by the inclusion of either pia mater or the central retinal vessel. They showed that including pia mater in the eye geometry resulted in an average relative difference of 26.9%, 18.2%, and 50.5% in the maximum principal strain estimated within the LC, prelaminar neural tissue, and postlaminar neural tissue, respectively. (Mao et al., 2020) constructed an axisymmetric generic eye geometry, and evaluated the maximum $1^{st}$ principal strain within the LC at different IOP values using finite element analysis. At an IOP of 26 mmHg, the tensile strain within the LC reached a maximum value of 6.25% for this eye model with simplified/generic geometry. In contrast, (Kharmyssov et al., 2021) found a much lower maximum tensile strain within the LC of approximately 3% using subject-specific finite element eye models based on *in vivo* imaging data with approximately the same IOP (25 mmHg). Inconsistency was also observed in terms of the effect of the translaminar pressure difference (TLPD), and more specifically ICP variations on the mechanics of eye tissues, particularly the LC (Feola et al., 2016a; Hua et al., 2017; Karimi, Razaghi, et al., 2022; Mao et al., 2020; Sigal et al., 2004). For example, (Hua et al., 2017) found a decrease up to 7.65% in the maximum $1^{st}$ principal strain within the LC when the ICP increased from 7 to 15 mmHg. (Feola et al., 2016a) also demonstrated that ICP elevation from 0 to 20 mmHg led to a decrease in the maximum tensile and compressive strains within the LC by 38.8% and 29.4%, respectively. Similarly, with a decrease of ICP from 8 to 2 mmHg, (Mao et al., 2020) observed an increase in both maximum tensile and compressive strains by 15% and 7%, respectively. Conversely, (Sigal et al., 2004) found an opposite trend to the aforementioned examples. More specifically, an elevation of cerebrospinal fluid pressure (or ICP) from 0 to 15 mmHg led to an increase in the average $1^{st}$ principal strain within the LC by 9.5% and 10.3% using their eye model with and without pia mater, respectively.

While multiple studies evaluated the mechanics of eye tissues, there are multiple variations in the modeling assumptions in these studies regarding the components of the eye that are necessary to its mechanical function, which has led to inconsistencies in the observations of the mechanical response of some eye tissues. Part of the reason for this variation in modeling assumptions is that it is not possible to differentiate or even identify some tissue regions from typical imaging data, and if included in a model, these regions must therefore be estimated based on standard dimensions from the literature. Overall, it is not yet clear what specific eye tissue regions need to be resolved in subject-specific eye geometries for robust representation of the eye, and particularly ONH mechanics. As such, the current study aims to establish the tissue regions necessary to accurately represent the mechanics of the ONH using subject-specific finite element eye geometries derived from *in vivo* human imaging data, and with a particular focus on those tissue regions that are not visible in the imaging data. This work then seeks to test this established modeling approach by evaluating the effect of pressure variations, particularly the IOP and ICP, on the mechanical response of the LC.

The following section outlines the methodology used to construct subject-specific finite element eye mechanical models, including details on geometry construction, material properties, and boundary conditions. The subsequent results section is divided into four subsections, with the first three subsections focused on establishing the tissue regions necessary in the eye geometry for an accurate representation of the mechanics of the ONH. The first subsection discusses the impact of changing the various eye tissues on the principal strains within four sectors of the LC using both local sensitivity analysis of material stiffness and a test with specific tissue regions entirely removed (i.e., binary tests of tissue regions existing in the model or not). Similarly, the second and third subsections present the sensitivity analysis and binary test results for the effect on the principal strains within the retina and the optic nerve, respectively. Finally, the fourth subsection presents the effect of IOP and ICP variations on the principal strains within the LC. This is followed by concluding remarks, summarizing the key findings, limitations, and implications of the present work.

## Methods and Materials

### 1. Data Acquisition

Two anonymized male subjects that were enrolled in a study conducted by Massachusetts Eye and Ear at Harvard Medical School were used as example cases to evaluate the consistency of the estimations of the mechanical behavior of the ONH (D. Li et al., 2017). Swept-source optical coherence tomography (SS-OCT) (Topcon DRI OCT-1 Atlantis; Topcon Inc, Tokyo, Japan; center wavelength: 1050 nm; scan speed: 100,000 A-scans per second) images were acquired for the right eye of both subjects. The SS-OCT scans covered an area of 6×6 mm² of the ONH with a lateral resolution of 20 μm and an axial resolution of 8 μm. Both cases were diagnosed as healthy subjects with no history of glaucoma. The subjects were aged 61 and 73 years old on the day of imaging and had right eye spherical equivalents of -0.75 and 2.625, respectively. Informed consent was obtained from the subjects enrolled in the study and the study protocol was in compliance with the Declaration of Helsinki and was approved by the Institutional Review Board (IRB) of Massachusetts Eye and Ear, Harvard Medical School (Protocol No. 2019P002755).

### 2. 3D Reconstruction of Subject-Specific Eye Geometry

Subject-specific eye geometries for each subject were created using 6 radial SS-OCT slices of the ONH spaced at 30° rotational intervals. Each of the 6 two-dimensional (2D) images was manually segmented under the supervision of two expert ophthalmologists to identify the tissue regions corresponding to the retina, choroid, sclera, LC, and optic nerve. However, as is typical for such imaging data, the posterior boundary of the LC was not visible in the images (Girard et al., 2015) and neither were the boundaries of several potentially important tissue regions, including the pia mater, dura mater, Bruch's membrane, border tissue, and annular ring. Therefore, the dimensions of these nonvisible regions were defined based on 'standard' geometric data from the literature (Chong et al., 2010; Curcio & Johnson, 2013; Feola et al., 2016a; Satekenova et al., 2019; Wang et al., 2016, 2018). As noted, the posterior limit of the sclera was captured manually with standard contrast and brightness enhancement of the images. The LC thickness for each subject was described to be 0.38 mm based on published data for healthy subjects (Jonas & Holbach, 2005). The thickness of pia mater was set to 0.06 mm (Satekenova et al., 2019; Wang et al., 2016, 2018), whereas dura mater had a thickness of 0.3 mm with an increased thickness at the peripapillary sclera insertion of 0.75 mm (Feola et al., 2016a; Wang et al., 2016). Lastly, the Bruch's membrane was modeled as a shell (i.e., a thin curved structure for which the midsurface defines the geometry) with a constant thickness of 3 μm (Chong et al., 2010; Curcio & Johnson, 2013), and was embedded in the eyeball between retina and choroid. The border tissue and the annular ring had a thickness of 0.06 mm and 0.45 mm, respectively (Wang et al., 2018). Figure 1 shows an example SS-OCT slice with each segmented/defined tissue region labeled for one of the subjects considered.

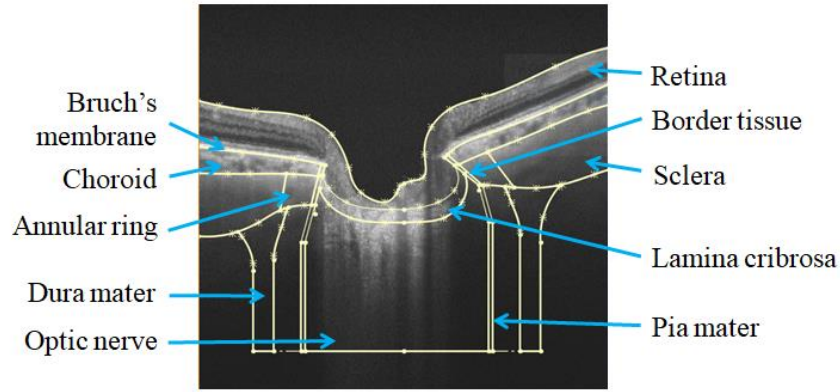

Figure 1: Example slice from the SS-OCT scan of the optic nerve head with the segmentation shown for all tissue regions evaluated for one of the subjects.

After each 2D image was segmented, an arc in the circumferential direction was created defining the interface between the ONH and the eye globe. Closed surfaces bounding each tissue region were then created by linearly interpolating each segmented slice edge in the anterior and posterior directions and interpolating along the arc circumferentially. Then, the closed surfaces representing each tissue region were converted into solid volumes and boundaries between these solids were merged to create a continuous solid comprised of the multiple tissue regions. The complete 3D ONH geometry was embedded within a hollow sphere representing the eye globe with estimated subject-specific axial length, with a constant thickness corresponding to the average combined thickness of the sclera, choroid, and retina over the 6 SS-OCT slices. More specifically, the most anterior point of the sphere, representing the cornea, was positioned based on the axial length of the subject. Then, circular arcs were created by connecting the top corneal point to a circle with the axial length as diameter at the equatorial plane and to the edges of each tissue from each radial slice within the ONH. Similarly to the ONH geometry, closed surfaces for each tissue regions within the sphere were created by circumferentially interpolating the arcs of the sphere along the curvature of the ONH as the guide curve. Again, the closed surfaces were converted to solid volumes and merged.

Axial length was not available as a direct measure for the subjects considered in this study. Therefore, the axial length values were estimated using the subject-specific spherical equivalent values based on the following empirical formula (Y.-Y. Chen et al., 2022):

$$AL = -0.389\ SE + 23.75, \qquad \text{Equation 1}$$

where AL refers to the axial length (mm) and SE is the spherical equivalent (diopter). The ONH was located at an offset position for both subjects, such that the horizontal angular distance was 15.5° between fovea and the center of the ONH (Rohrschneider, 2004). Figure 2 shows an example of the complete 3D subject-specific eye geometry for one of the subjects. The segmentation and 3D reconstruction of the subject-specific eye geometries were carried out using Solidworks software (Dassault Systemes Solidworks Corp., 2021).

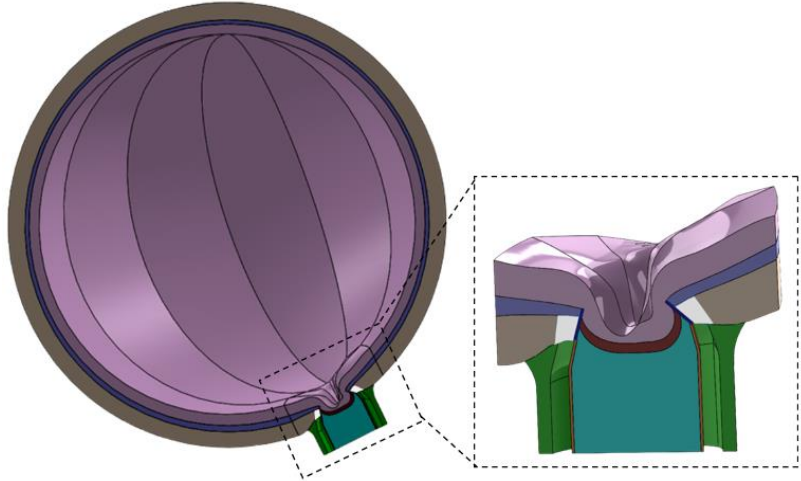

Figure 2: Reconstructed 3D subject-specific eye geometry

## 3. Eye Mechanical Analysis

Assuming all applied forces and other boundary conditions were sufficiently constant to neglect any dynamic effects, standard static finite element analysis was applied to estimate the subject-specific mechanical response (i.e., displacement, stress, and strain) of the eyes due to variations in IOP and ICP.

All tissue regions detailed in the prior section were assumed to be fully bonded to each other at all interfaces. In addition, support conditions were applied to prevent rigid body motion. More specifically, boundary conditions were applied to four points located at the equator of the eye globe (two points each at the median and horizontal plane of the eye), to prevent all motions tangential to the surface of the eye globe. Then, the IOP and ICP were applied as constant uniform pressures to the inner surface of the eye globe and to the subarachnoid space (i.e., surface between dura mater and pia mater), respectively. The magnitudes of both IOP and ICP were assumed to be in the normal range of a healthy adult, with IOP of 20 mmHg and ICP of 10 mmHg (Munakomi & M Das, 2024; Ren et al., 2010), for all tests to assess the sensitivity of the ONH mechanical response to the material properties of the tissues. Then, for the final set of tests, the magnitudes of both pressures were varied to evaluate the effect of changes in IOP and ICP on the mechanical response of the eyes.

Similar to other studies on ocular biomechanics, all tissue regions were assumed to be homogeneous within the designated region (i.e., the complete eye was considered discretely heterogenous), and all tissues were assumed to be isotropic and nearly incompressible in response to the applied IOP and ICP, except for retina and optic nerve that were considered partially compressible (Chuangsuwanich et al., 2020; Jafari et al., 2021; Kharmyssov et al., 2021; Sigal et al., 2012). Although these noted studies similarly assumed isotropy, it should be acknowledged that fibrous soft tissues, such as sclera and LC, are known to potentially exhibit anisotropy depending on the loading conditions. Yet, such behavior should not have an effect on the overall goal of the present study, which is to evaluate the relative effect/importance of tissue regions, and not to establish precise subject-specific properties for the subjects considered. To confirm this assumption, preliminary tests were performed including arbitrary levels of anisotropy in the tissue regions, which did not have a significant effect on the overall impact of the tissue material properties on the biomechanical response observed within the ONH (results not shown for brevity). In terms of the specific material models, based on similar recent research (Feola et al., 2018; Jin et al., 2018; Wang et al., 2018), the Bruch's membrane, choroid, retina, optic nerve, and border tissue were modeled as linear elastic materials. Alternatively, it has been noted that sclera, LC, pia mater, dura mater, and annular ring can exhibit significant nonlinearity, and were therefore treated as hyperelastic Neo-Hookean materials (Chuangsuwanich et al., 2020; Feola et al., 2016a, 2018; Wang et al., 2016). The constitutive models with the baseline mechanical properties of each eye tissue are summarized in Table 1. It should be noted that the parameters for the annular ring, LC, and sclera were based only on those reported for the ground matrix stiffness of the tissue, again based on the previously noted ability to neglect anisotropy for the present study. Moreover, although the LC and sclera have been modeled with a hyperelastic Mooney Rivlin model previously (Wang et al., 2016), when only considering the ground matrix, these models simplify to Neo-Hookean, which was used herein.

Table 1: The constitutive models and baseline material parameters for each eye tissue region, including the Young's modulus (E) and Poisson's ratio (ν).

| Tissue | Constitutive Model | Material Properties | Source |
|---|---|---|---|
| Sclera | Neo-Hookean | E = 4.8 MPa, $\nu = 0.49$ | (Wang et al., 2016) |
| LC | Neo-Hookean | E = 0.3 MPa, $\nu = 0.49$ | (Wang et al., 2016) |
| Bruch's membrane | Elastic | E = 10.79 MPa, $\nu = 0.49$ | (Jin et al., 2018) |
| Choroid | Elastic | E = 0.6 MPa, $\nu = 0.49$ | (Wang et al., 2018) |
| Retina | Elastic | E = 0.05 MPa, $\nu = 0.45$ | (Feola et al., 2016a) |
| Optic nerve | Elastic | E = 0.05 MPa, $\nu = 0.45$ | (Feola et al., 2016a) |
| Border tissue | Elastic | E = 10.79 MPa, $\nu = 0.49$ | (Jin et al., 2018) |
| Dura mater | Neo-Hookean | E = 3 MPa, $\nu = 0.49$ | (Chuangsuwanich et al., 2020) |
| Pia mater | Neo-Hookean | E = 3 MPa, $\nu = 0.49$ | (Chuangsuwanich et al., 2020) |
| Annular ring | Neo-Hookean | E = 0.596 MPa, $\nu = 0.49$ | (Feola et al., 2018) |

All mechanical analyses were performed using the commercial finite element analysis software ABAQUS (Dassault Systemes, 2021) with quadratic tetrahedron continuum elements for all tissue regions other than the Bruch's membrane, which had triangular 3-node general purpose shell elements. Standard mesh convergence was performed to ensure a sufficiently small mesh size was used for accurate simulations in all test cases. A total of 253,710 elements were deemed sufficient for the analysis of the subject shown.

## Results and Discussion

To establish the tissue regions necessary to accurately represent the mechanics of the ONH, a first set of tests was performed to evaluate the sensitivity of each region of interest (i.e., LC, retina, and optic nerve) with respect to the material properties of various ONH tissues using two subject-specific eye finite element models. For these local sensitivity analyses, the Young's modulus of each tissue was increased and decreased by 25% from the baseline value given in Table 1. It should be noted that these property variations are still within the ranges reported in the literature for healthy subjects. Then, a follow-on set of tests was conducted to assess the effect of removing entirely the tissues with the lowest effect on the responses of the regions of interest to confirm whether there was added value or not to their inclusion. Lastly, as an example application utilizing the results from the first two sets of tests, the subject-specific finite element modeling with the established sensitive and necessary tissues was used to evaluate the effect of IOP and ICP variations on the deformation (i.e., strain) response of the LC.

For all test cases, the compressive (i.e., directional minimum) and tensile (i.e., directional maximum) principal strain values of each region were extracted and evaluated in terms of their largest magnitude (i.e., peak value) and average magnitude within the specified region. It should be noted that the peak tensile and compressive strain values were computed as the 95$^{th}$ percentile and 5$^{th}$ percentile of the extracted principal strain values, respectively, to avoid numerical artifacts due to shape irregularity (Feola et al., 2016a). For the local sensitivity analysis, the relative difference was calculated between each regional average and peak strain component estimated with both increased and decreased tissue moduli with respect to the strains estimated with the baseline moduli. Then, the overall impact from the moduli variations was quantified as the average of the relative differences from increasing and decreasing the moduli values. Additionally, according to previous studies, regional structural differences have been observed within the LC in particular (e.g., LC pore size) (Kiumehr et al., 2012), which assists in understanding certain diseases, such as the association of glaucoma with localized neural loss (A. J. Tatham et al., 2013). Therefore, the regional behavior of the estimated mechanical responses of the LC was evaluated in more detail by dividing the LC into four subregions: temporal, inferior, nasal, and superior, as shown in Figure 4. To further confirm the level of the mechanical effect, changes in the average direction of the maximum principal strains of each region were also calculated. It should be noted that results from only one of the two subjects are presented since, although the deformation responses were somewhat different due to the changes in subject geometry, all significant observations were consistent for both subjects.

As an initial point of reference, the table within Figure 3 shows the average and peak tensile and compressive strain values within the LC subregions, the retina, and the optic nerve for one subject with the baseline material properties for all tissue regions. Figure 3 also shows the tensile and compressive strain distributions within the ONH at an IOP of 20 mmHg and an ICP of 10 mmHg. One particular observation is that there are higher compressive and tensile strains within both the retina and the optic nerve compared to the LC. This can be explained by the fact that the retina is directly subjected to IOP in contrast to the LC, while the optic nerve is more affected by the ICP, which is transmitted through the pia mater. There are also higher strains within the temporal and inferior LC subregions compared with the other two subregions, which is consistent with those observed by previous studies (Beotra et al., 2018; D. E. Midgett et al., 2017).

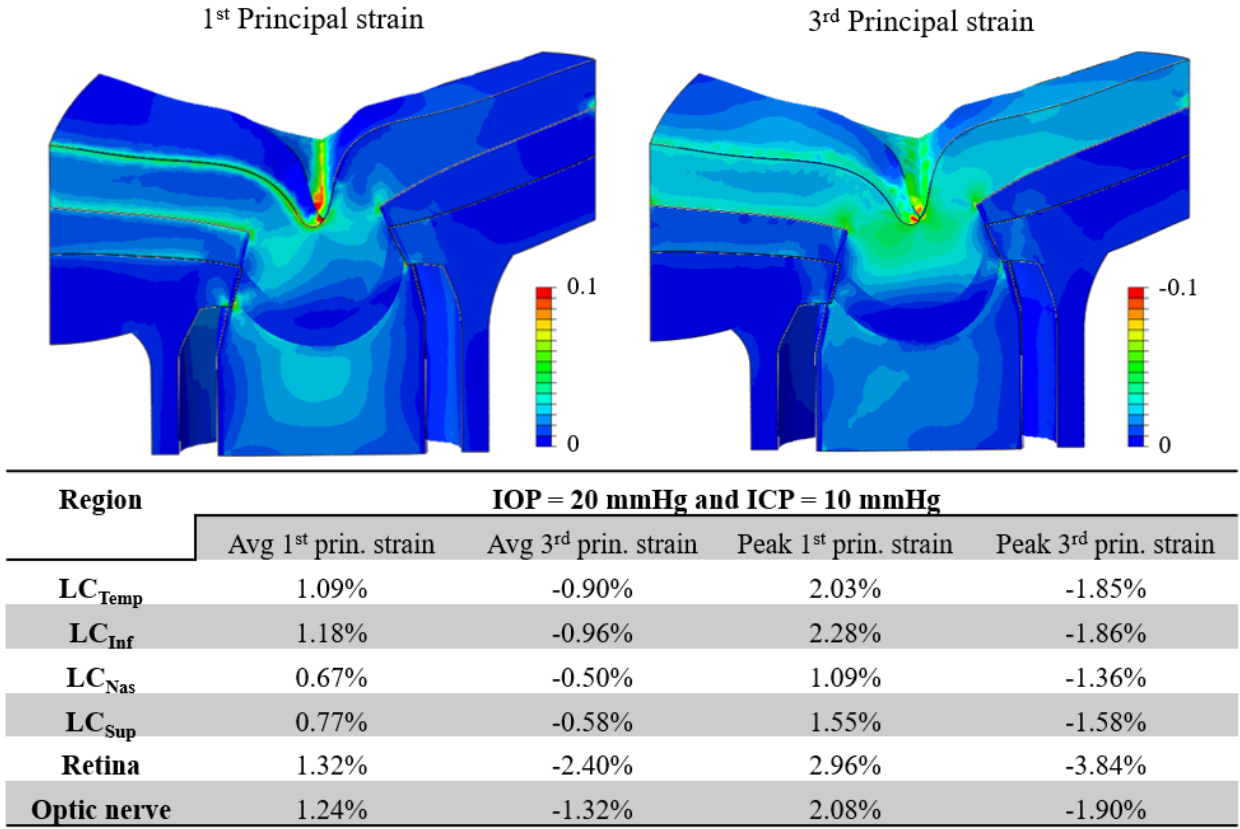

| Region | IOP = 20 mmHg and ICP = 10 mmHg | | | |
|---|---|---|---|---|
| | Avg 1$^{st}$ prin. strain | Avg 3$^{rd}$ prin. strain | Peak 1$^{st}$ prin. strain | Peak 3$^{rd}$ prin. strain |
| $LC_{Temp}$ | 1.09% | -0.90% | 2.03% | -1.85% |
| $LC_{Inf}$ | 1.18% | -0.96% | 2.28% | -1.86% |
| $LC_{Nas}$ | 0.67% | -0.50% | 1.09% | -1.36% |
| $LC_{Sup}$ | 0.77% | -0.58% | 1.55% | -1.58% |
| Retina | 1.32% | -2.40% | 2.96% | -3.84% |
| Optic nerve | 1.24% | -1.32% | 2.08% | -1.90% |

Figure 3: 1st and 3rd average principal strains within the LC subregions, retina, and the optic nerve (numerical results) and the tensile and compressive strain distributions within the ONH at an IOP of 20 mmHg and an ICP of 10 mmHg (color contours).

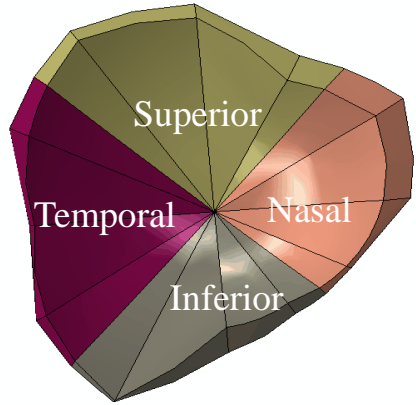

Figure 4: The subject-specific LC divided into four subregions: superior, nasal, inferior, and temporal.

## 1. Effect of Materials on Lamina Cribrosa Mechanics

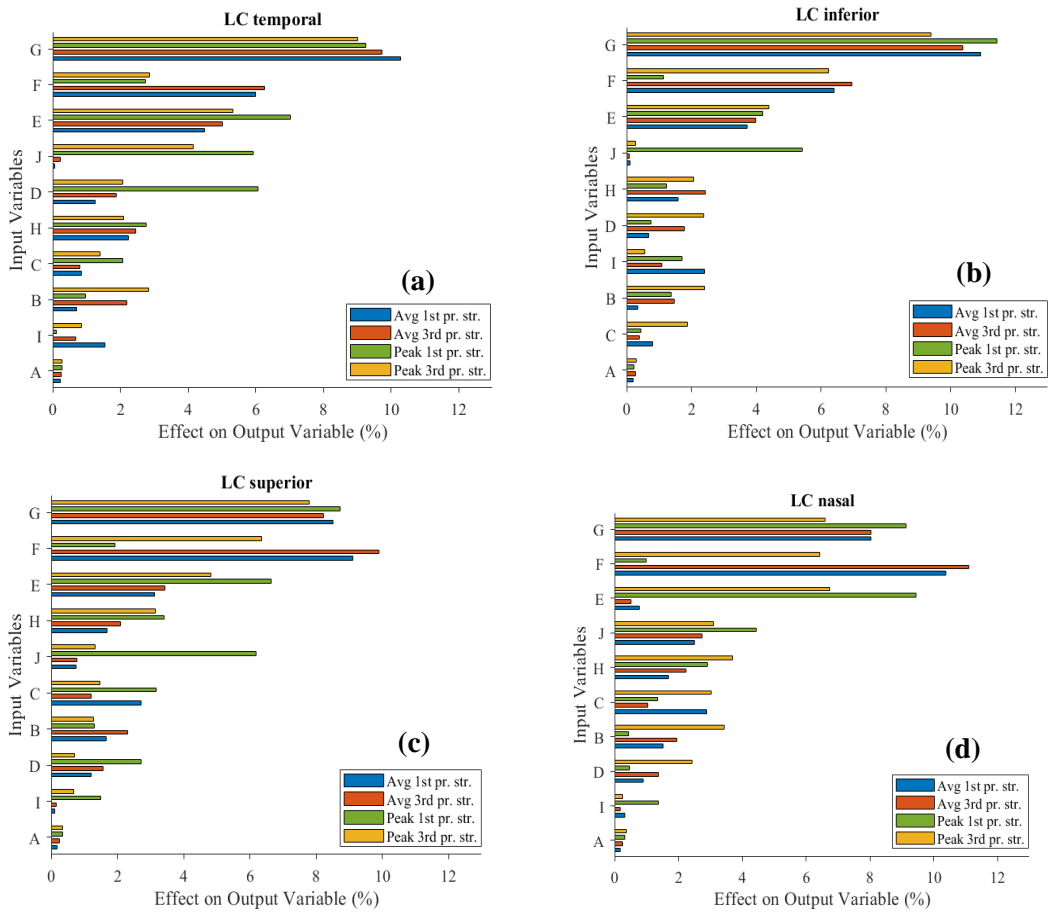

Figure 5: Sensitivity analysis of the effect of regional ONH tissue properties on the tensile (1st pr. Str.) and compressive (3rd pr. Str.) strains in terms of spatial average (Avg) and maximum (Peak) within the a) temporal, b) inferior, c) superior, and d) nasal subregions of the LC. The regional properties varied

correspond to:  A- Bruch's membrane, B- border tissue, C- pia mater, D- annular ring, E- dura mater, F- LC, G- sclera, H- choroid, I- retina, J- optic nerve.

Figure 5 shows the impact of each tissue's modulus variation on the spatial average and peak compressive and tensile strains within the four subregions of the LC. An important first observation is that no significant differences are seen between the effect on the compressive versus the tensile strains for most tissues. Although retinal elastic modulus was found to affect the laminar tensile strains more than the compressive strains within all laminar subregions. It was also found that varying the elastic modulus of the optic nerve significantly affected the laminar maximum tensile strains in most LC subregions, and had a lower effect on the compressive strains. This variable effect can be explained by the type of loading that the tissue is transmitting to the LC. The retina is directly transmitting the IOP to the LC, and since the IOP has the effect of 'inflating' the eye, which primarily 'stretches' the surrounding tissues (H. Tran et al., 2018), changes in retina primarily result in changes to the tensile strains in the LC. Alternatively, the optic nerve being positioned below the LC, provides significant support to the LC in resisting the deformation due to the IOP as well. More compliant optic nerve leads to higher tension and lower compression within the LC, with a more significant effect on the tensile strains.

Overall, the four tissues that were found to have the largest effect on the principal strains within all LC subregions were: sclera, LC, dura mater, and optic nerve, with total averaged impacts of both tension and compression, peak, and spatial average values over all subregions of 9.1%, 5.9%, 4.3%, and 3.0%, respectively. These tissues also had a large effect on the average principal direction within the LC due to the material property variations. The stiffness of the sclera had the largest overall effect on both the tensile and compressive strains within the majority of the LC, even more of an effect than the stiffness of the LC itself. This significance of the sclera can be explained by the structural support that the sclera provides to the ONH as the outer most layer. Consequently, changes in scleral material properties lead to an altered biomechanical behavior of the overall ONH, including the LC. These findings are consistent with other studies in mechanics of the ONH (Eilaghi et al., 2010; Wang et al., 2016). For instance, (Sigal et al., 2004) reported that the sclera acts as a "protective shield" for the LC and that variations in sclera material properties had the greatest effect on the magnitude of LC deformation. Furthermore, (Coudrillier et al., 2016) demonstrated that the stiffening of the sclera contributes to a reduction in pressure-induced deformations within the LC. The dura mater is the tissue that has the next largest effect on the mechanical response of the LC, other than the LC itself. The dura mater, being one of the stiffest tissues in the ONH, has a significant role in transmitting the ICP to its surrounding tissues, the sclera and pia mater. The stresses due to the ICP are transmitted to sclera from dura mater and then transmitted to the LC, significantly affecting strains within the LC. Consistent findings were reported in the literature by (Hua et al., 2018), who classified dura mater elastic modulus as one of the most influential factors on the biomechanics of the ONH. Similar to dura mater, the optic nerve had a considerable effect on strains within the LC subregions. The optic nerve is particularly significant as it both transmits the stresses due to ICP to the LC and provides foundational support to the LC to resist deformation due to IOP.

The stiffness of the Bruch's membrane, retina, pia mater, border tissue, and annular ring had the least effect (i.e. an effect of 2% or less, which is more than an order of magnitude less than the change in material parameter value for each test) on the mechanical response of the LC, with total averaged impacts over the subregions of 0.3%, 0.8%, 1.6%, 1.6%, and 1.8%, respectively. These tissues also exhibited the least effect on the average principal direction within the LC due to the material property variations. The pia mater, which is relatively stiff compared to the surrounding tissues, contributes to transmitting the stresses due to the ICP to the optic nerve, which is 60 times more compliant than the pia mater (Sigal et al., 2005). The compliance of the optic nerve allows it to deform and accommodate the mechanical stresses transmitted by the pia mater, thereby reducing the effect of the pia mater stiffness on the strain distribution within the LC. Similar results were found by (Sigal et al., 2004) and (Hua et al., 2018) supporting the weak dependence of the LC deformation on pia mater stiffness. The annular ring is located between the sclera and the border tissue, which are both stiffer than the annular ring. Therefore, changes in stiffness of the annular ring had a limited influence on the forces transmitted from the annular ring to either the border tissue or the sclera, and ultimately the strains within the LC. Lastly, even though Bruch's membrane and border tissue are two of the stiffest tissues in the ONH, they are also the thinnest tissues in the ONH and represent a relatively small amount of the total tissue volume of the ONH, which leads to their limited effect on the mechanics of the LC. Finally, the retina also had a limited impact on the strains within the LC subregions, which is likely due to the compliant nature of the tissue. The retina is over 10 times more compliant than its neighboring tissues, and given that it is directly subjected to the IOP, the retina deforms significantly and less force is transferred to adjacent tissues.

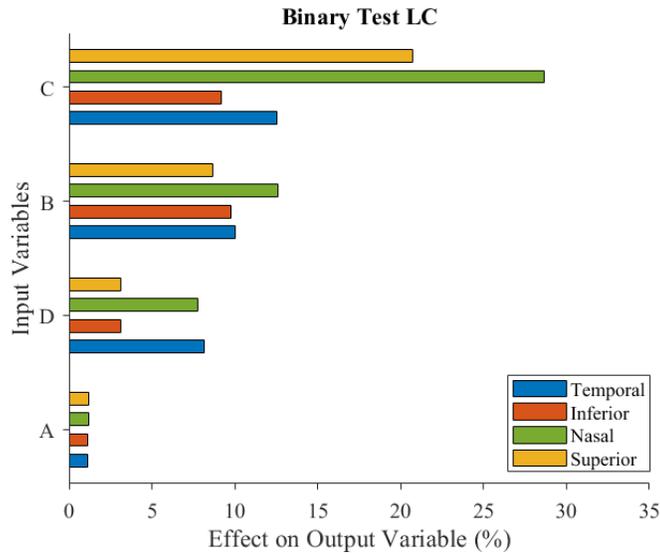

Figure 6: Binary test effect on strains within the LC subregions. A- Bruch's membrane, B- border tissue, C- pia mater, D- annular ring.

The follow-on test was performed to evaluate the effect of removing entirely the Bruch's membrane, pia mater, border tissue, and annular ring, in turn (i.e., one at a time), from the subject-specific finite element models. Specifically, the Bruch's membrane shell geometry embedded between the retina and the choroid was entirely eliminated from the subject-specific models, but with all other tissues as originally described with the baseline Young's modulus values. Whereas the border tissue, annular ring, and pia mater were assigned the material properties of the regions they were adjacent to, thereby extending the adjacent tissue region to replace that tissue. Although retina properties had one of the least effect on strains within the LC, the tissue was not further tested in the binary test. The complete removal of retina will make the LC directly subjected to the IOP, changing the boundary conditions and significantly affecting the LC deformation. Also, neglecting the properties of the retina by extending the choroid, which is 10 times stiffer than retina will also generate significant change in stresses transmitted to the LC. The limited effect of retina properties on the strains within the LC suggests that the choice of "accurate" retina material parameters is not critical to the mechanical response of the LC, provided the retina stiffness is within the range estimated herein. However, the inclusion/existence of retina within the eye geometry is necessary. Figure 6 shows the effect of the removal of Bruch's membrane, pia mater, border tissue, and annular ring on principal strains within the four subregions of the LC. These results are presented as a single average of the relative change in the compressive and tensile strains for each LC subregion. The existence of both the border tissue and pia mater result in a more than 10% change in LC strains on average for all four subregions, with the largest changes within the nasal and superior subregions. The existence of the pia mater is significant since it is 60 times stiffer than the adjacent optic nerve tissue. Therefore, the replacement of the pia mater with additional optic nerve tissue resulted in the ICP deforming the entire region of the optic nerve more than when the pia mater was present, ultimately leading to additional compression on the lower boundary of the LC. In short, the pia mater 'shields' the LC from some amount of the ICP. For the border tissue, the upper border tissue was replaced by extending the choroid and the lower border tissue was replaced by extending the annular ring. Both the choroid and annular ring are more than 15 times more compliant than the border tissue. Therefore, the removal of the border tissue had a considerable effect on the forces of both IOP and ICP transmitted to the LC, leading to a significant effect on the strain levels within the LC. The existence of the annular ring had an impact of approximately 8% on the temporal and nasal LC subregion strains, but had a significantly lower effect on the other subregions. In particular, the LC was found to deform less in the temporal and nasal subregion when the annular ring was included in contrast to when it was not, since the relatively soft annular ring deforms significantly and transfers less force to the LC. The temporal side is likely the most affected subregion due to its location being farthest from the boundary conditions (i.e., the equator), compared with the other subregions. Alternatively, the existence of the Bruch's membrane had a nearly negligible effect on the strain values within the LC, with an average change of approximately 1.2% in strains averaged between tension and compression and over the four subregions. This minimal effect is likely due to the distance of the Bruch's membrane from the LC and the small thickness of the Bruch's membrane. In summary, the inclusion of the annular ring, the border tissue, and pia mater is important to evaluating the mechanics of the LC, even though their respective material property values do not have a significant effect on the response. However, the Bruch's membrane is likely not required to be included to accurately estimate the mechanical response of the LC.

## 2. Effect of Materials on Retina Mechanics

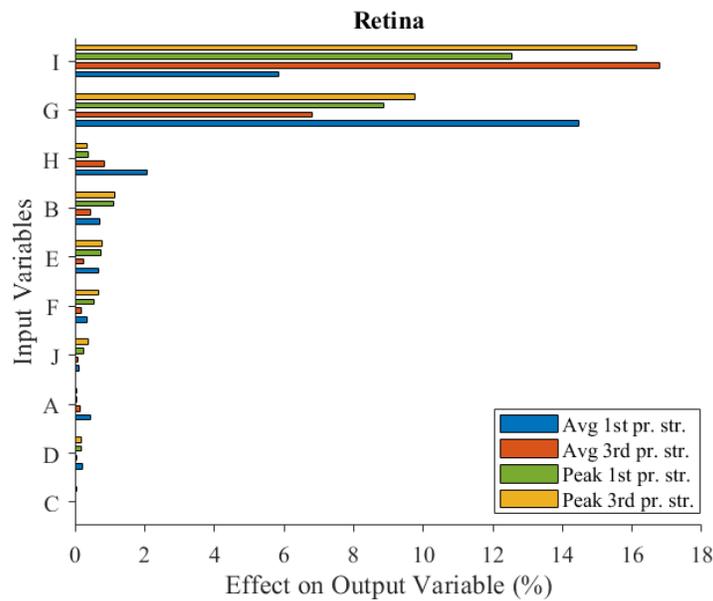

Figure 7: Sensitivity analysis of the effect of regional ONH tissue properties on the tensile (1$^{st}$ pr. Str.) and compressive (3$^{rd}$ pr. Str.) strains in terms of spatial average (Avg) and maximum (Peak) within the retina. The regional properties varied correspond to: A- Bruch's membrane, B- border tissue, C- pia mater, D- annular ring, E- dura mater, F- LC, G- sclera, H- choroid, I- retina, J- optic nerve.

Figure 7 shows the impact of each tissue's elastic modulus variation on the spatial average and peak compressive and tensile strains within the retina. Only two tissues had an impact greater than 2% on the principal strains within the retina, which were the retina itself and the sclera. Unlike the LC, the stiffness of the retina itself has the largest effect on the strains experienced within the retinal tissue, with a total averaged impact of both tension and compression, peak, and spatial average values of 12.8%. Additionally, the stiffness of the sclera has a substantial effect on the retina deformation, with a total averaged change in strains of 10.0%, further emphasizing its overall structural importance to eye tissues. Similarly, the retina and sclera properties also had the largest effects on the average principal strain directions within the retina. All tissues except for the retina and sclera had a limited effect (i.e. less than 2%) on the strains within the retina. Since the retina is directly subjected to the IOP, a change in the material properties of the surrounding tissues has a relatively low effect on the retinal mechanical response. The material properties of the pia mater, annular ring, Bruch's membrane, optic nerve, LC, dura mater, border tissue, and choroid had total averaged changes in strains within retina of 0.02%, 0.2%, 0.2%, 0.2%, 0.4%, 0.6%, 0.8%, and 0.9%, respectively. Also, these tissues had the least effect on the average principal direction within retina. It should be noted though that the change in the choroid and border tissue has a more significant effect on the tensile response compared to the compressive response. Reducing the stiffness of sclera and choroid allows for more expansion (i.e., stretching) of the retina, and thus, more significantly increases the average retinal tensile strains than the compressive strains, with respective increases of 14.5% and 2.1% in tensile strains compared with only 6.8% and 0.9% in compressive strains.

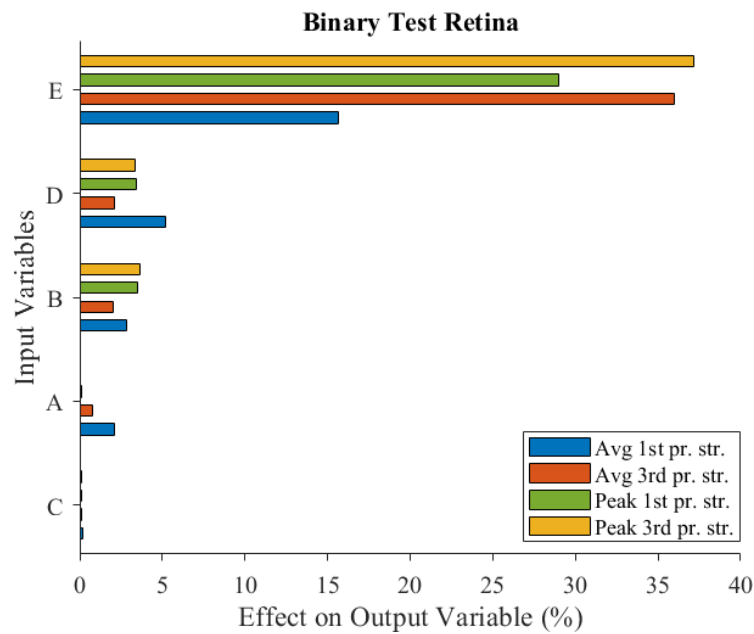

Figure 8: Binary test effect on the tensile (1$^{st}$ pr. Str.) and compressive (3$^{rd}$ pr. Str.) strains in terms of spatial average (Avg) and maximum (Peak) within the retina. A- Bruch's membrane, C- pia mater, D- annular ring, E- dura mater.

Due to the low sensitivity of the retina mechanical response to changes in their elastic moduli, the effect of removing entirely the border tissue, Bruch's membrane, pia mater, annular ring, and dura mater was tested (Figure 8). The border tissue, Bruch's membrane, pia mater, and annular ring were removed with the same method as in Section 1. The removal of the dura mater was simulated by completely removing the tissue region from the analysis. Similar to the retina noted in the previous Section, the removal of the choroid, LC, and optic nerve were not tested due to their important roles in maintaining the structural support and geometric representation of the ONH. Although, noting their lack of associated response sensitivities, "accurate" parameter estimates for these tissues would not be needed for evaluating the retina mechanics. The removal of the dura mater, annular ring, border tissue, Bruch's membrane, and pia mater had impacts on the retina deformation, with total averaged impacts of both tension and compression, peak, and spatial average values of 29.4%, 3.5%, 3.0%, 0.7%, and 0.1%, respectively. The large effect of removing the dura mater on strains within retina is likely due its role in providing support to the overall tissues within the ONH. More specifically, the removal of dura mater allowed for significant increase in tension within all tissues in the ONH, including the retina. Also, the removal of dura mater led to an increase in compressive strains experienced by the sclera due to the ICP, thus affecting the compression levels within the retina as well. Removing the border tissue by extending both the choroid and the annular ring which are both compliant compared with the border tissue, resulted in transmission of less force from the ICP to the retina. This minimal effect on retinal strains can be attributed to the low thickness of the border tissue, and the different forces experienced by both tissues. The retina is directly subjected to the IOP, whereas the border tissue is mostly affected by the ICP. Removing the annular ring also has a limited impact on retinal strains, but there is a more significant change in the tensile strains compared to the compressive strains. Extending the stiffer sclera tissue to remove/replace the annular ring provides more structural support to the eye overall, reducing the total amount of stretch due to the IOP, therefore more significantly effecting the tensile strains. Although it is directly connected to the retina, the entire removal of Bruch's membrane still had a relatively minor effect on both the compressive and tensile strains within the retina. Similar to the LC effect, this lack of sensitivity is likely due to the small thickness of the Bruch's membrane. However, similar to the annular ring, Bruch's membrane affects the overall resistance to the IOP, and thus affects tensile strains more compared with compressive strains. The low sensitivity to the existence of pia mater is likely due to their relatively large distance from the retina. This low sensitivity is likely further due to the distinct biomechanical forces experienced by the retina (primarily influenced by the IOP) compared to the pia mater (subjected to the ICP). More specifically, the removal of the pia mater results in a direct transmission of forces onto the optic nerve, which is more likely to affect the surrounding tissues, such as the LC. Thus, the more distant retina remains unaffected by this alteration. These findings emphasize the limited impact of the border tissue, Bruch's membrane, annular ring, and pia mater on the mechanical response of the retina.

### 3. Effect of Materials on Optic Nerve Mechanics

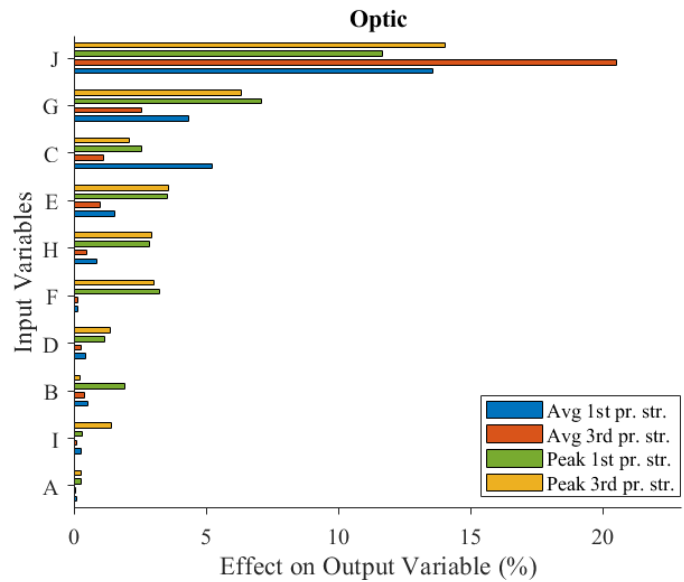

Figure 9: Sensitivity analysis of the effect of regional ONH tissues properties on the tensile (1$^{st}$ pr. str.) and compressive (3$^{rd}$ pr. str.) strains in terms of spatial average (Avg) and maximum (Peak) within the optic nerve. The regional properties varied correspond to: A- Bruch's membrane, B- border tissue, C- pia mater, D- annular ring, E- dura mater, F- LC, G- sclera, H- choroid, I- retina, J- optic nerve

The impact of the elastic modulus variations of the ONH tissues on the spatial average and peak compressive and tensile strains within the optic nerve is shown in Figure 9. The four tissues with the highest effect on strains within the optic nerve were the optic nerve, sclera, pia mater, and dura mater. More specifically, the optic nerve itself exhibited the highest effect, with a total averaged impact of both tension and compression, peak, and spatial average values of 14.9%. The surrounding sclera, known for its structural support, also demonstrated a relatively important impact on optic nerve strains with a total averaged impact of 5.1%. Furthermore, the pia mater, responsible for transmitting forces of the ICP to the optic nerve, emerged as another influential tissue, with a total averaged impact of 2.7%. The dura mater, characterized by its high stiffness, had a total averaged impact of 2.4%, which is likely due to the change in ICP forces transmitted to the pia mater, thus indirectly influencing the forces transferred from the pia mater to the optic nerve. The material properties of the optic nerve, pia mater, sclera, and dura mater also had the highest effect on the average principal strain directions within the optic nerve. In contrast, the Bruch's membrane, retina, annular ring, border tissue, LC, and choroid exhibited negligible effects (i.e. less than 2%) on the optic nerve strains, with total averaged impacts of 0.2%, 0.5%, 0.8%, 0.8%, 1.6%, and 1.8%, respectively. Similarly, the material parameters of these tissues also had the least effect on the average principal directions within the optic nerve.

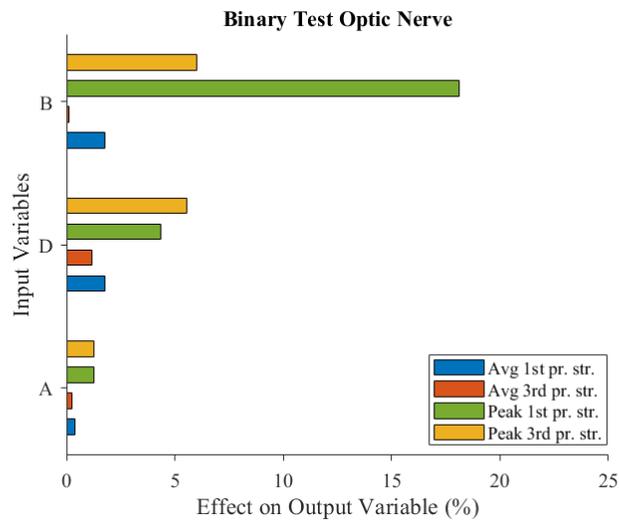

Figure 10: Binary test effect on the tensile (1$^{st}$ pr. str.) and compressive (3$^{rd}$ pr. str.) strains in terms of spatial average (Avg) and maximum (Peak) within the optic nerve. A- Bruch's membrane, B- border tissue, D- annular ring.

The effect of the removal of the Bruch's membrane, the border tissue, and the annular ring on the compressive and tensile strains within the optic nerve is shown in Figure 10. All tissues were removed with the same procedure described in Section 1. Again, despite the low impact of retina, LC, and choroid properties on strains within the optic nerve, the existence of these tissues in the eye geometry is necessary. Therefore, these tissues were not further tested in the binary test. The existence of the border tissue was found to have a significant effect, with a total averaged impact of both tension and compression, peak, and spatial average values of 6.5%, and with a particularly large effect on the peak principal strain values within the optic nerve (both tension and compression exceeding 10% change). Although the border tissue elastic modulus value did not show a significant influence on optic nerve strains, replacing this tissue by extending the 10 times compliant choroid and annular ring tissue decreased the amount of force from both IOP and ICP transmitted to the pia mater. Thus, the force transmitted from the pia mater to the optic nerve also decreased. The removal of the annular ring and Bruch's membrane had minimal effects, with total averaged impacts of 3.2% and 0.8%, respectively. Despite the high stiffness of Bruch's membrane within the ONH, this tissue has a minimal influence on the optic nerve due to its low thickness and distance from the optic nerve. The existence of the annular ring also had a limited impact, which could be due to its anatomical position between two relatively stiff tissues (i.e. sclera and border tissue), thus making the effect of its removal on strains within the optic nerve limited.

## 4. Effect of Pressure Variations on Lamina Cribrosa Mechanics

Changes in IOP and ICP have been noted to lead to alterations in the biomechanical behavior of the LC, which in many cases has been investigated due to the hypothesized link between the mechanics of the LC and the state or progression of glaucoma (D. Midgett et al., 2020; Price et al., 2020; Roberts et al., 2010b). However, as noted in the Introduction, there have been some significant variations in the modeling strategies used to evaluate LC mechanics and some of the core observations relating to the effects of IOP and ICP on the biomechanical behavior of the LC. As such, the results of the sensitivity testing herein, particularly those for evaluating the LC in Section 1, were applied to again create subject specific eye models to evaluate the effects of variations in both IOP and ICP. Furthermore, the results of this evaluation are compared with observations of similar studies from the literature.

According to the results from Section 1, the two subject specific eye mechanical models were again created following the same procedure as previously, but in this case including only the tissues that were significant to the mechanics of the LC, which resulted in the exclusion of the Bruch's membrane. Three different values of IOP were applied, in turn: 10, 20, and 30 mmHg, and for each IOP value, three values of ICP were applied: 5, 10, and 15 mmHg.

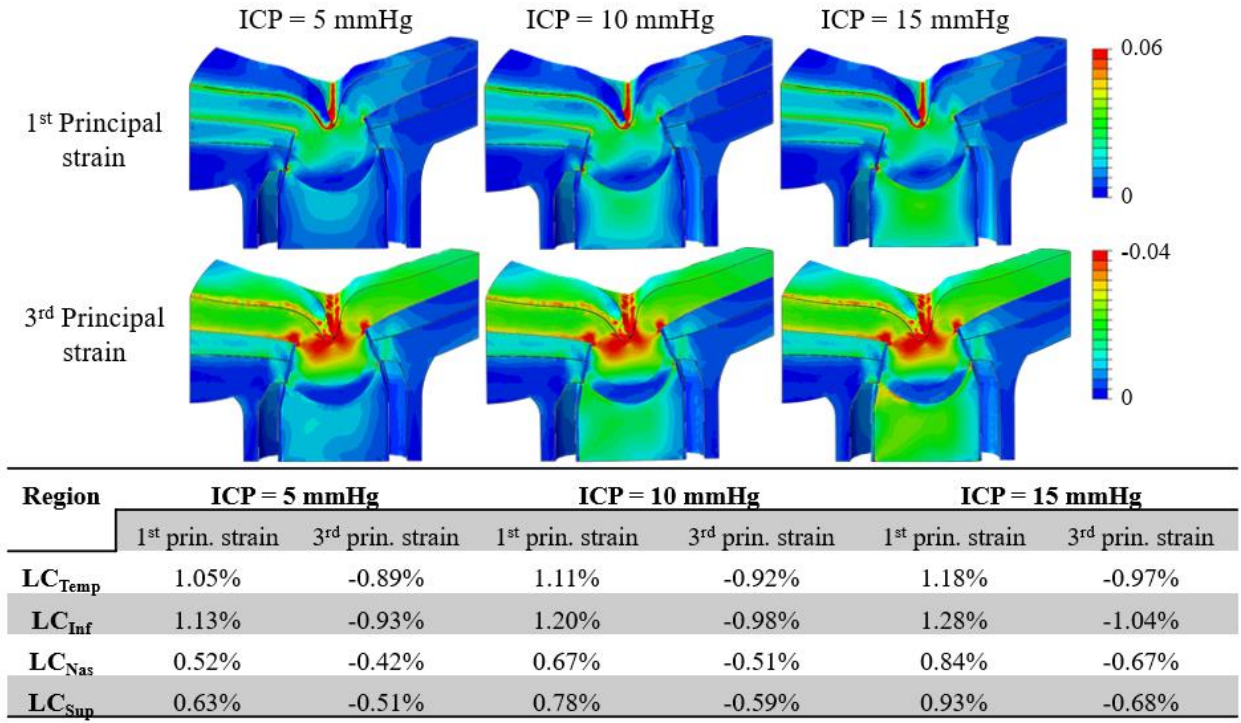

| Region | ICP = 5 mmHg | | ICP = 10 mmHg | | ICP = 15 mmHg | |
|---|---|---|---|---|---|---|
| | 1st prin. strain | 3rd prin. strain | 1st prin. strain | 3rd prin. strain | 1st prin. strain | 3rd prin. strain |
| $LC_{Temp}$ | 1.05% | -0.89% | 1.11% | -0.92% | 1.18% | -0.97% |
| $LC_{Inf}$ | 1.13% | -0.93% | 1.20% | -0.98% | 1.28% | -1.04% |
| $LC_{Nas}$ | 0.52% | -0.42% | 0.67% | -0.51% | 0.84% | -0.67% |
| $LC_{Sup}$ | 0.63% | -0.51% | 0.78% | -0.59% | 0.93% | -0.68% |

Figure 11: 1st and 3rd average principal strains within the LC subregions averaged over the three IOP values (numerical results) and the tensile and compressive strain distributions within the ONH at the IOP of 20 mmHg (color contours) at ICP values of 5, 10, and 15 mmHg.

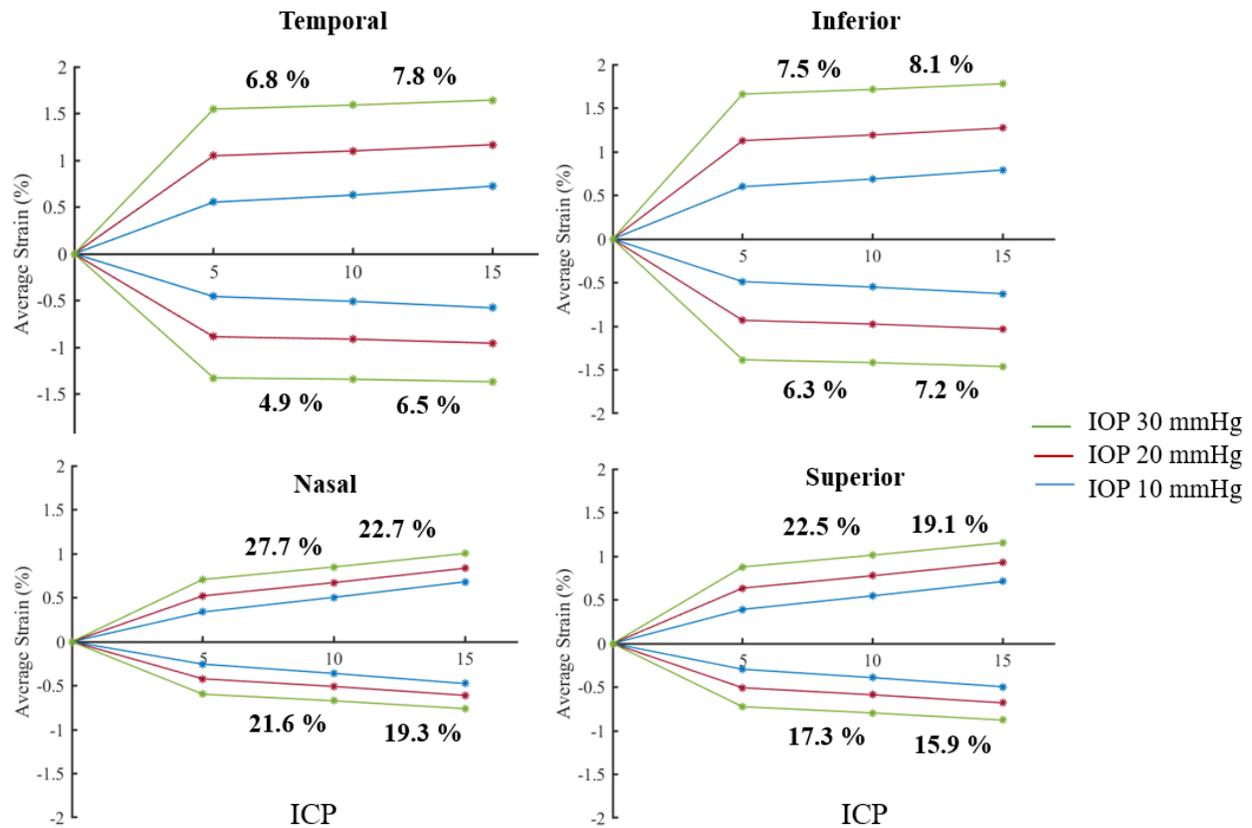

Figure 12: The compressive and tensile strain values at different values of IOP and ICP within a) the temporal, b) the inferior, c) the nasal, and d) the superior subregion of the LC.

The table within Figure 11 shows the average compressive and tensile strain values within the LC subregions, averaged over the three values of the IOP, for each value of the ICP. Figure 11 also shows the compressive and tensile strain distributions within the ONH at an IOP of 20 mmHg for each value of the ICP. It should be noted that the scale of the tensile and compressive strain values have been adjusted to highlight the strain distribution within the LC. The strain distribution shows an increase in ONH deformation, particularly within the optic nerve and the LC, when the ICP increases for both tensile and compressive strains. These outcomes appear dependent in that elevated ICP leads to higher stresses within the optic nerve, consequently increasing stresses transmitted to the LC. More specifically, the tensile strains, averaged over the LC subregions, increased from 0.83% to 0.94% to 1.06% when the ICP was increased from 5 mmHg to 10 mmHg then to 15 mmHg. Similarly, the compressive strains, averaged over the LC subregions, also had an increase from -0.69% to -0.75% to -0.84% with the same variation of the ICP. Figure 12 shows the spatial average of the compressive and tensile principal strains within the four subregions of the LC with respect to the applied IOP and ICP. Increasing IOP and ICP values led to elevated strain values within the LC. Specifically, the temporal and inferior sectors exhibited an increase of tensile and compressive principal strain values of more than 6% and more than 4%, respectively, averaged for all values of IOP for each increase of 5 mmHg of the ICP. The nasal and superior sectors experienced significantly higher increases in principal strains, with tensile and compressive principal strain values of more than 19% and more than 15%, respectively, averaged for all values of IOP for each increase of 5 mmHg of the ICP. It was also observed that the average tensile and compressive strain levels were higher at the temporal and inferior subregion at all pressure cases, compared with the other two subregions. More specifically, the compressive and tensile strains within temporal and inferior subregion were found to be almost 2 times higher than strains within the nasal and superior subregions.

The present findings align with multiple studies that observed increased LC tensile and compressive strains in response to elevated cerebrospinal fluid pressure, which increases the ICP. For instance, (Feola et al., 2017) experimentally found similar results, observing an increase in LC tensile strains from approximately 2.4% to 3.3%, and an increase in compressive strains from around -1.9% to -2.9%, when the ICP was increased from 10 to 20 mmHg. While the relative change in the experimental results is higher than those observed herein of 0.9% to 1.1% and -0.75% to -0.84%, respectively, averaged over the LC subregions and over the three IOP values when the ICP increased from 10 mmHg to 15 mmHg (Figure 11), the trend is consistent, especially for the tensile strains. Moreover, some variation is always expected, particularly considering that the experimental study was *ex vivo* and the current study utilized *in vivo* data. In particular, it has been noted that computational models often underestimate tissue strains in comparison with experimental studies (Feola et al., 2017). The observations herein are also consistent with other studies that observed high strain levels within the temporal and inferior subregions of the LC due to increases in IOP. For instance, using human donor LCs with no history of glaucoma, (D. E. Midgett et al., 2017) found higher tensile strain values in the inferior quadrant and the temporal quadrant compared to the other LC regions when the IOP was increased to 45 mmHg. An *in vivo* study of healthy eyes also showed higher strains in the inferotemporal region of the LC following an increase of IOP to 35 mmHg (Beotra et al., 2018).

Although the findings in the present study were consistent with several reported results from previous studies, there are other studies that have reported contradictory results. For example, (Feola et al., 2016b; Jóhannesson et al., 2018; Mao et al., 2020) observed an increase in LC strains with decreasing ICP. For example, using a generic (i.e., non-subject-specific) finite element eye model, (Mao et al., 2020) showed an increase in strains within the LC with an increase in TLPD due to either an increase of IOP or a decrease of ICP. Similarly, (Feola et al., 2016b) reported a decrease in LC tensile and compressive strains with increased ICP values using a generic finite element eye model. However, it should be noted that these seemingly contradictory findings may be attributed to various differences in subjects and/or modeling assumptions used. Specifically, the effect of TLPD appears more pronounced when the LC is thinner (Mcmonnies, 2016), which is also associated with an increased translaminar pressure gradient (i.e., (IOP–ICP) / LC thickness). Also, tests with animal models showed different impacts of ICP variations on the ONH (Feola et al., 2017). Overall, the variability of findings, such as these effect of ICP on LC mechanics further emphasizes the need for a consistent modeling approach, as evaluated herein, to more precisely and consistently evaluate the influence of factors, such as ICP on the mechanical behavior of the eye.

## Conclusions and Future Directions

A computational study was used to evaluate the effect of various ONH tissues on the mechanical response of the pathologically significant regions of the eye. In particular, the effect of varying material properties and/or the inclusion of ONH tissues on the strain fields within the LC, the retina, and the optic nerve was evaluated. Two healthy subjects were considered for *in vivo* computational mechanical analysis, but all results were consistent between the two subjects, therefore only results of one subject were shown. After evaluating the effects of the various ONH tissues, the tissues identified as important to the eye mechanical response were included in a final set of subject-specific eye models and used to evaluate the effect of pressure variations on the mechanical response of the LC. The results showed that the stiffness of the sclera has one of the largest impacts on the mechanical behavior of the eye, particularly the LC. Other than their own respective properties, the properties of the dura mater and pia mater also significantly affect the eye regions of interest, specifically the LC and optic nerve, respectively. In contrast, the Bruch's membrane had little effect on the mechanics of the LC. Therefore, the Bruch's membrane is likely not required to be included to accurately estimate the mechanical response of the LC. Similarly, the border tissue, Bruch's membrane, annular ring, and pia mater all had limited impact on the mechanical response of the retina. Lastly, the Bruch's membrane and the annular ring had minimal influence on the mechanical response of the optic nerve. The pressure variations evaluated using the modeling approach defined by the previous tests were consistent with examples in the published literature, e.g., showing an increase in strain values within the LC when both the IOP and the ICP increased, with the highest strains localized at the inferior and temporal subregions of the LC, and providing confidence in the validity of the method. However, there is also some disagreement in such observations in the literature, further emphasizing the need for deeper investigations on biomechanical processes in the eye, such as the effect of translaminar pressure difference on the LC. As such, this study has accomplished an important step in establishing an eye modeling process for more consistent assessment of ocular mechanics, specifically highlighting the necessary ONH components to accurately evaluate various pathologically relevant tissue regions, such as the mechanical response of the LC for potential application in glaucoma research. Extending the present research towards such future work would likely require improving physiological accuracy and subject-specificity. In particular, future studies to evaluate subject-specific variations in ONH mechanics should likely include the anisotropy of the soft tissue components to more accurately assess the spatial distribution of the mechanical response. Although, an additional challenge may be to determine the subject-specific *in vivo* properties for such models, and potentially even additional heterogeneity of the properties, to see the full benefit of their inclusion. Similarly, subject-specific geometry of eye models could be further improved, such as better representation of the eye globe when additional imaging data is available.

**Data availability statement**

All data, models, or code that support the findings of this study are available from the corresponding author upon reasonable request.

**Ethical approval**

The study protocol was in compliance with the Declaration of Helsinki and was approved by the Institutional Review Board (IRB) of Massachusetts Eye and Ear, Harvard Medical School (Protocol No. 2019P002755).

**Declaration of competing interests**

The authors declare that they have no known competing financial interests or personal relationships that could have appeared to influence the work reported in this paper.

**Funding**

This research did not receive any specific grant from funding agencies in the public, commercial, or not-for-profit sectors.

McKean-Cowdin, R., Varma, R., Wu, J., Hays, R. D., & Azen, S. P. (2007). Severity of Visual Field Loss and Health-related Quality of Life. *American Journal of Ophthalmology*, *143*(6), 1013–1023. https://doi.org/10.1016/j.ajo.2007.02.022

McKean-Cowdin, R., Wang, Y., Wu, J., Stanley, P. A., & Rohit, V. (2008). *Impact of Visual Field Loss on Health-Related Quality of Life in Glaucoma: The Los Angeles Latino Eye Study—ScienceDirect*.

Mcmonnies, C. W. (2016). The interaction between intracranial pressure, intraocular pressure and lamina cribrosal compression in glaucoma. *Clinical and Experimental Optometry*, *99*(3), 219–226. https://doi.org/10.1111/cxo.12333

Midgett, D. E., Pease, M. E., Jefferys, J. L., Patel, M., Franck, C., Quigley, H. A., & Nguyen, Thao. D. (2017). The Pressure-Induced Deformation Response of the Human Lamina Cribrosa: Analysis of Regional Variations. *Acta Biomaterialia*, *53*, 123–139. https://doi.org/10.1016/j.actbio.2016.12.054

Midgett, D., Liu, B., Ling, Y. T. T., Jefferys, J. L., Quigley, H. A., & Nguyen, T. D. (2020). The Effects of Glaucoma on the Pressure-Induced Strain Response of the Human Lamina Cribrosa. *Investigative Ophthalmology & Visual Science*, *61*(4), 41. https://doi.org/10.1167/iovs.61.4.41

Munakomi, S., & M Das, J. (2024). Intracranial Pressure Monitoring. In *StatPearls*. StatPearls Publishing. http://www.ncbi.nlm.nih.gov/books/NBK542298/

Muñoz-Sarmiento, D. M., Rodríguez-Montaño, Ó. L., Alarcón-Castiblanco, J. D., Gamboa-Márquez, M. A., Corredor-Gómez, J. P., & Cortés-Rodríguez, C. J. (2019). A finite element study of posterior eye biomechanics: The influence of intraocular and cerebrospinal pressure on the optic nerve head, peripapillary region, subarachnoid space and meninges. *Informatics in Medicine Unlocked*, *15*, 100185. https://doi.org/10.1016/j.imu.2019.100185

Nishinaka, A., Inoue, Y., Fuma, S., Hida, Y., Nakamura, S., Shimazawa, M., & Hara, H. (2018). Pathophysiological Role of VEGF on Retinal Edema and Nonperfused Areas in Mouse Eyes With Retinal Vein Occlusion. *Investigative Ophthalmology & Visual Science*, *59*(11), 4701–4713. https://doi.org/10.1167/iovs.18-23994

Phillips, J. D., Hwang, E. S., Morgan, D. J., Creveling, C. J., & Coats, B. (2022). Structure and mechanics of the vitreoretinal interface. *Journal of the Mechanical Behavior of Biomedical Materials*, *134*, 105399. https://doi.org/10.1016/j.jmbbm.2022.105399

Price, D. A., Harris, A., Siesky, B., & Mathew, S. (2020). The Influence of Translaminar Pressure Gradient and Intracranial Pressure in Glaucoma: A Review. *Journal of Glaucoma*, *29*(2), 141–146. https://doi.org/10.1097/IJG.0000000000001421

Ren, R., Jonas, J. B., Tian, G., Zhen, Y., Ma, K., Li, S., Wang, H., Li, B., Zhang, X., & Wang, N. (2010). Cerebrospinal fluid pressure in glaucoma: A prospective study. *Ophthalmology*, *117*(2), 259–266. https://doi.org/10.1016/j.ophtha.2009.06.058

Roberts, M. D., Sigal, I. A., Liang, Y., Burgoyne, C. F., & Downs, J. C. (2010a). Changes in the Biomechanical Response of the Optic Nerve Head in Early Experimental Glaucoma. *Investigative Opthalmology & Visual Science*, *51*(11), 5675. https://doi.org/10.1167/iovs.10-5411

Roberts, M. D., Sigal, I. A., Liang, Y., Burgoyne, C. F., & Downs, J. C. (2010b). Changes in the Biomechanical Response of the Optic Nerve Head in Early Experimental Glaucoma. *Investigative Ophthalmology & Visual Science*, *51*(11), 5675–5684. https://doi.org/10.1167/iovs.10-5411

Rodriguez-Beato, F. Y., & De Jesus, O. (2023). Compressive Optic Neuropathy. In *StatPearls*. StatPearls Publishing. http://www.ncbi.nlm.nih.gov/books/NBK560583/

Rohrschneider, K. (2004). Determination of the Location of the Fovea on the Fundus. *Investigative Ophthalmology & Visual Science*, *45*(9), 3257–3258. https://doi.org/10.1167/iovs.03-1157

Satekenova, E., Ko, M. W. L., & Kim, J. R. (2019). Investigation of the Optic Nerve Head Morphology Influence to the Optic Nerve Head Biomechanics – Patient Specific Model. *2019 41st Annual International Conference of the IEEE Engineering in Medicine and Biology Society (EMBC)*, 5370–5373. https://doi.org/10.1109/EMBC.2019.8856743
22